\documentclass[aps,prl,twocolumn,superscriptaddress]{revtex4}

\usepackage{mathrsfs,amssymb,graphics,subfigure,threeparttable}
\usepackage{grap hicx}
\usepackage{amssymb}
\usepackage{enumerate}
\usepackage{amsmath}
\usepackage{fancyhdr}
\usepackage{bm}
\usepackage[squaren]{SIunits}

\begin{document}


\title{Phase Space Dynamics of Ionization Injection in Plasma Based Accelerators}


\author{X. L. Xu}
\affiliation{Key Laboratory of Particle and Radiation Imaging of Ministry of Education, Tsinghua University, Beijing 100084, China}
\author{J. F.  Hua}
\affiliation{Key Laboratory of Particle and Radiation Imaging of Ministry of Education, Tsinghua University, Beijing 100084, China}
\author{F. Li}
\affiliation{Key Laboratory of Particle and Radiation Imaging of Ministry of Education, Tsinghua University, Beijing 100084, China}
\author{C. J. Zhang}
\affiliation{Key Laboratory of Particle and Radiation Imaging of Ministry of Education, Tsinghua University, Beijing 100084, China}
\author{L. X. Yan}
\affiliation{Key Laboratory of Particle and Radiation Imaging of Ministry of Education, Tsinghua University, Beijing 100084, China}
\author{Y. C. Du}
\affiliation{Key Laboratory of Particle and Radiation Imaging of Ministry of Education, Tsinghua University, Beijing 100084, China}
\author{W. H. Huang}
\affiliation{Key Laboratory of Particle and Radiation Imaging of Ministry of Education, Tsinghua University, Beijing 100084, China}
\author{H. B. Chen}
\affiliation{Key Laboratory of Particle and Radiation Imaging of Ministry of Education, Tsinghua University, Beijing 100084, China}
\author{C. X. Tang}
\affiliation{Key Laboratory of Particle and Radiation Imaging of Ministry of Education, Tsinghua University, Beijing 100084, China}
\author{W. Lu}
\email[]{weilu@tsinghua.edu.cn}
\affiliation{Key Laboratory of Particle and Radiation Imaging of Ministry of Education, Tsinghua University, Beijing 100084, China}
\affiliation{University of California Los Angeles, LA, CA 90095, USA}
\author{P. Yu}
\affiliation{University of California Los Angeles, LA, CA 90095, USA}
\author{W. An}
\affiliation{University of California Los Angeles, LA, CA 90095, USA}
\author{W. B. Mori}
\affiliation{University of California Los Angeles, LA, CA 90095, USA}
\author{C. Joshi}
\affiliation{University of California Los Angeles, LA, CA 90095, USA}

\date{\today}

\begin{abstract}
The evolution of beam phase space in ionization-induced injection into plasma wakefields is studied using theory and particle-in-cell (PIC) simulations. The injection process causes special longitudinal and transverse phase mixing leading initially to a rapid emittance growth followed by oscillation, decay, and eventual slow growth to saturation. An analytic theory for this evolution is presented that includes the effects of injection distance (time), acceleration distance, wakefield structure, and nonlinear space charge forces. Formulas for the emittance in the low and high space charge regimes are presented. The theory is verified through PIC simulations and a good agreement is obtained. This work shows how ultra-low emittance beams can be produced using ionization-induced injection.
\end{abstract}


\maketitle

The field of plasma based acceleration has experienced significant progress in the past decade \cite{RevModPhys.81.1229}. $\giga\electronvolt$ energy gain in centimeter-scale laser driven wakes (LWFA) has been achieved in many recent experiments \cite{leemans2006gev, PhysRevLett.103.035002, PhysRevLett.103.215006, PhysRevLett.105.105003}. In beam driven wakes (PWFA), high gradient acceleration has been sustained over meter-scale distances leading to more than $40\giga\electronvolt$ energy gain \cite{PhysRevLett.93.014802, PhysRevLett.95.054802, blumenfeld2007energy}. For future applications of wakefield accelerators such as FELs and colliders,  the quality of the self-injected beams in plasma waves, namely the transverse and longitudinal emittances, need to be improved and controlled.  Among the many injection schemes \cite{PhysRevLett.102.065001, PhysRevLett.100.215004}, ionization-induced injection methods have attracted significant interests due to its simplest and flexibility \cite{PhysRevLett.98.084801, PhysRevLett.100.105005, PhysRevLett.104.025003, PhysRevLett.104.025004,  PhysRevLett.105.105003, PhysRevLett.107.035001, PhysRevLett.107.045001}. However, the injection process involves complex phase space dynamics, and the achievable final beam quality strongly depends on this evolution process. This area of research is of fundamental importance for achieving  beam quality well beyond what is achievable with current technology.  
In this letter, we examine carefully the effects that affect the beam phase space evolution in ionization-induced injection using a combination of theory and simulations. We found the evolution typically has three stages, and each stage can impact  the final beam quality. In typical cases where the injection time is limited to few inverse plasma periods ($2\pi\omega_p^{-1}$) and the charge is low, the three stages are as follows.  First, when ionization is occurring,  the emittance of the injected beam grows quickly in time from the initial thermal emittance. Second, immediately following ionization, the emittance slowly decreases to a minimum value. Finally, the emittance again gradually increases to saturated values. If the ionization time is more than $\sim\pi \omega_p^{-1}$ then the emittance grows to the saturated level during the first stage including  an oscillatory behavior before it slowly decreases. In the ``high" charge limit the emittance evolves monotonically towards the same saturated value.

The theory reveals that the evolution in emittance described above is due to special longitudinal and transverse phase mixing of electrons born at different times. The derived expressions clearly show how the emittance depends on different physical parameters, e.g., injection distance, acceleration distance, energy spread, wakefield structure and  nonlinear self-forces. The predictions are compared against results from OSIRIS PIC simulations \cite{fonseca2002high} and good agreement is obtained.

\begin{figure}[bp]
\includegraphics[width=0.5\textwidth]{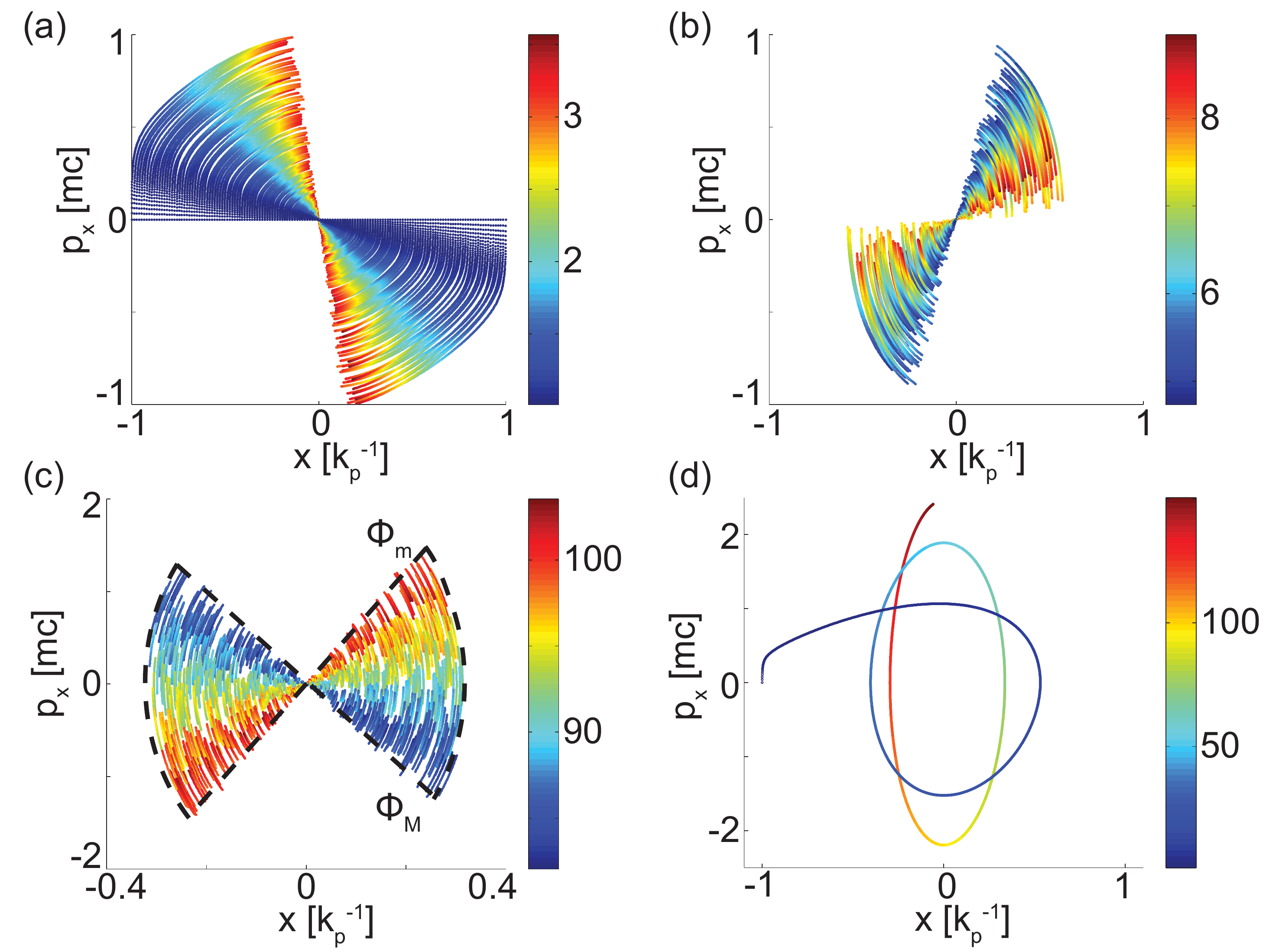}
\caption{\label{fig1} Snapshots from single particle simulations illustrating the transverse phase mixing. Snapshots (a) to (c) show the $x-p_x$ phase space when (a) t=3 and the injection is terminated at this time, (b) t=8, (c) t=94.  (d) shows the $x-p_x$ phase space trajectory for a particle. The color corresponds to the relativistic factor $\gamma$ which at early times is correlated with the ionization time, $s_i$.}
 \end{figure}
 
To understand the emittance evolution observed in numerous simulations, we first analyze the single particle motion with the equation of motion for a charged particle undergoing acceleration and betatron motion \cite{PhysRevLett.88.135004} in a perfectly linear focusing force, such as in a nonlinear  wake excited in the blowout regime \cite{PhysRevLett.96.165002}\cite{lu2006nonlinearPoP},
\begin{align}
\ddot{\vec{x}}_{\perp} + \frac{\dot{\gamma}}{\gamma}\dot{\vec{x}}_{\perp} + k_\beta^2 \vec{x}_{\perp} = 0 \label{motion equation}
\end{align}
where $\cdot$ refers to derivative with respect to $s=z$, the propagation distance, $\vec{x}_\perp$ is the transverse coordinates of the particle, $k_\beta = k_p/\sqrt{2\gamma}$ is the betatron wave number, $k_p \equiv \omega_p / c$. Here the variable $ct-z \equiv \xi$ is introduced to define the longitudinal position of the particle inside the wake. In the limit that $\ddot{\gamma} \ll k_\beta \dot{\gamma}$ and $\dot{\gamma} \ll k_\beta \gamma$, Eq. (\ref{motion equation}) has a general asymptotic solution of the form $\vec{x}_\perp = \left(\vec{x}_{\perp 0} / \gamma^{1/4}\right) e^{i \int ds k_\beta}$. In addition exact solutions for Eq. (\ref{motion equation}) can be found for specific cases such as when $\dot{\gamma}=\mathrm{constant} = qE_z / m c^2$, which is reasonable when phase slippage is not important.  For this case Eq. (\ref{motion equation}) becomes,
\begin{align}
\ddot{x} + \frac{E_z}{\gamma_0 + E_z s} \dot{x} + \frac{x}{2\left( \gamma_0 + E_z s\right)} = 0 \label{motion equation constant Ez}
\end{align}
where $\gamma(s) = \gamma_0+E_z s$, and we normalize position to $k_p^{-1}$, time to $\omega_p^{-1}$, $E_z$ to $mc\omega_p / e$, and charge to $e$. Exact and asymptotic solutions for Eq. (\ref{motion equation constant Ez}) are,
\begin{align}
x &= c_1 J_0\biggl(\sqrt{\frac{2\gamma}{E_z^2}} \biggl) + c_2 Y_0\biggl( \sqrt{\frac{2\gamma}{E_z^2}}\biggl) \approx C \biggl( \frac{2E_z^2}{\pi^2 \gamma}\biggl)^{\frac{1}{4}} \mathrm{cos} \Phi   \label{eq3}\\
\dot{x} &= -\sqrt{\frac{1}{2\gamma}} \biggl[ c_1 J_1\biggl( \sqrt{\frac{2\gamma}{E_z^2}} \biggl) + c_2 Y_1\biggl( \sqrt{\frac{2\gamma}{E_z^2}}\biggl) \biggl] \nonumber \\
&\approx - C  \biggl( \frac{E_z^2}{2\pi^2 \gamma^3}\biggl)^{\frac{1}{4}} \mathrm{sin}\Phi \label{eq4}
\end{align}
where $\Phi = \left(\sqrt{2\gamma} - \sqrt{2\gamma_0}\right) / E_z$ is the betatron phase. The asymptotic solutions are of the general form with $\Phi = \int ds/\sqrt{2\gamma}$. Direct comparison with single particle and PIC simulations shows that the asymptotic expressions are valid with high accuracy when $\gamma \gtrsim 2$ \cite{Xu2013prst1}.

We consider the $x-p_x$ phase space corresponding to one of the two transverse directions. Electrons are defined by their ionization time, $s_i$, and initial phase space location $\left(x_0(s_i), p_{x0}(s_i)\right)$. In addition electrons of interest are rapidly accelerated as they reach a longitudinal position $\xi_f$ in the wake where they remain phase locked and thus feel a constant $E_z$. As we show later electrons ionized at the same $s_i$ can reside over the full range of $\xi_f$ within the bunch (we call this longitudinal phase mixing), and thus feel a range of $E_z$ which we define as $\delta E_z$. We also assume that each electron begins at rest and the rapid interaction with the incoming laser leads to a small natural spread in $p_{x}$. This ``thermal" spread can thus be neglected.   

We integrate Eq. (\ref{motion equation constant Ez}) for many test electrons. To model the effect of $\delta E_z$, the $E_z$ of each electron is randomly chosen from $0.9$ to $1.1$. In Fig. \ref{fig1}(a) we show electrons ionized at different times. After a group of electrons is ionized they begin to rotate in $x-p_x$ phase space. The first group (red) has a betatron phase, $\Phi_M$, and the most recent group (purple) has a phase, $\Phi_m$. If injection continues then $\Phi_m=0$. Clearly, the area in phase space increases during the injection process due to each group of electrons having a different betatron phase, we call this transverse phase mixing.

In Fig. \ref{fig1}(b) we show the phase space at a `time' after the injection has stopped. For simplicity we consider a case where the injection time, $\Delta s < \pi$. However electrons at $\Phi_M$ have a higher energy (due to being accelerated for a longer time) and hence lower betatron frequency than those electrons at $\Phi_m$. As a result $\Phi_M - \Phi_m\equiv \Delta \Phi$  gradually decreases and hence the emittance gradually decreases. 
 
Later in time, due to any spread in $E_z$ the electrons ionized at the same time develop a spread in phase. Eventually, the electrons at $\Phi_M$ ($\Phi_m$) are those injected first (last) but which have experienced the smallest (largest) $E_z$. In this case electrons at $\Phi_M$ now rotate faster than those at $\Phi_m$ causing the emittance to gradually increase.

\begin{figure}[bp]
\includegraphics[width=0.5\textwidth]{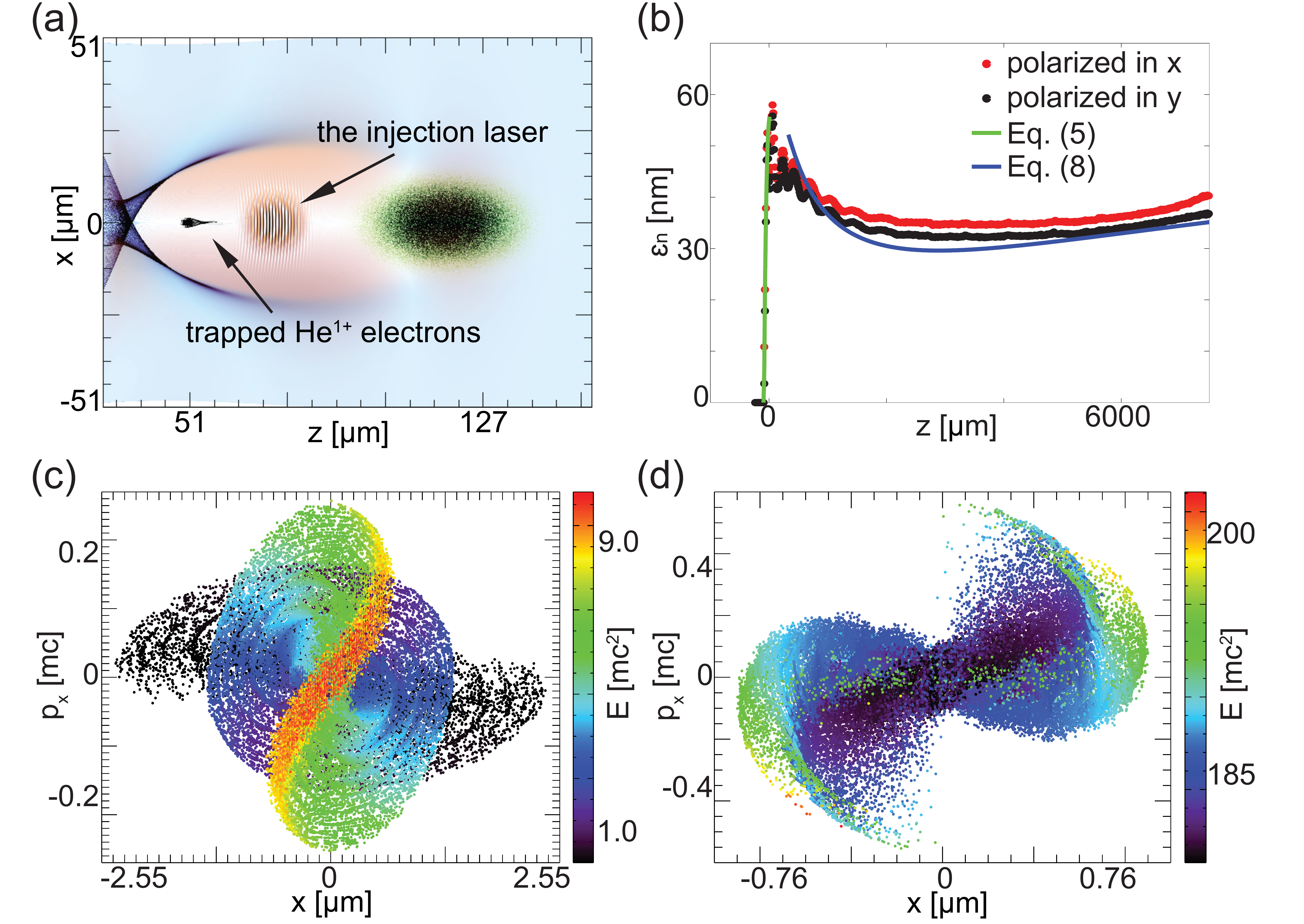}
\caption{\label{fig2} (a) 2D simulation of  laser injection into a beam driven wakefield.  Drive beam (green): $\sigma_r=15\micro\meter,\sigma_z=25\micro\meter,n_b=2.6\times 10^{17} \centi\meter^{-3},E_b=2\giga\electronvolt$. Laser: $\lambda=800\nano\meter, a_0=0.04, w_0=3\micro\meter, \tau \approx 30\femto\second$.  (b) Comparison of the emittance evolution between simulations and theory. The red and black lines are for the laser polarized in or out of the  simulation plane respectively. The $x-p_x$ phase space  when the injection is terminated (c) and  when $z=z_0$ (d).}
\end{figure}

It turns out a simple expression for the above emittance evolution process can be obtained if one assumes that the phase space distribution is independent of $\phi$ in a sector as for example shown by the dotted lines in Fig. \ref{fig1}(c). From Eqs. (\ref{eq3}) and (\ref{eq4}), one can see that the phase space coordinates $(x, p_x)$ depend very weakly on particle energy $(\sim \gamma^{1/4})$, therefore for a given time, we can assume $x=x_0 \left( \gamma_0 / \bar{\gamma} \right)^{1/4}\mathrm{cos}\Phi, p_x=x_0 \left( \gamma_0\bar{\gamma}\right)^{1/4}\mathrm{sin}\Phi/\sqrt{2}$, where the $\bar{\gamma}$ is the average energy of the injected particles at this time. We can obtain
 $\left<x^2\right>=\sigma_{x0}^2 (\gamma_0 / \bar{\gamma})^{1/2} [ 1 + (\mathrm{sin}2\Phi_M-\mathrm{sin}2\Phi_m) / (2\Phi_M-2\Phi_m)]  / 2$,  $\left<p_x^2\right>=\sigma_{x0}^2 (\gamma_0 \bar{\gamma})^{1/2} [1-  (\mathrm{sin}2\Phi_M-\mathrm{sin}2\Phi_m) / (2\Phi_M-2\Phi_m) ]/4$ and  $\left<xp_x\right>=( \sigma_{x0}^2 \gamma_0^{1/2}) (\mathrm{cos}2\Phi_M-\mathrm{cos}2\Phi_m) / (\Phi_M-\Phi_m)/4\sqrt{2}$,
where $\sigma_{x0}^2=\int x_0^2 f(x_0)dx_0 $ and $f(x_0)$ is the normalized distribution function when the electrons are born. Therefore the normalized emittance
\begin{align}
\epsilon_N\left( \Delta \Phi \right) = \sqrt{ \left< x^2 \right> \left< p_x^2\right> - \left< x p_x \right>^2} = \epsilon_{sat}\sqrt{1-\left(\frac{\mathrm{sin} \Delta \Phi}{\Delta \Phi}\right)^2} \label{eq5}
\end{align}
where $\Delta \Phi \approx \sqrt{2( E_{zm} z_M +1)} / E_{zm} - \sqrt{2( E_{zM} z_m +1)}/ E_{zM} $, $z_{M,m}=s-s_{iM,m}$, and $s_{iM,m}$ refers to when electrons at $M,m$ were ionized, $\epsilon_{sat}=\sigma_{x0}^2/2\sqrt{2}$ is the value of the emittance when the phase ellipse is filled out, in real units,
\begin{align}
\epsilon_{sat}[\micro\meter]=\frac{1}{2\sqrt{2}}k_p[\micro\meter^{-1}]\sigma_{x0}^2[\micro\meter^2] \label{eq6}
\end{align}
If we neglect  $\delta E_z$ which is reasonable early in time and assume injection is still occurring ($z_m=0$) and $\Delta \Phi  <1$, then $\epsilon_N \approx \epsilon_{sat}\sqrt{1/3} \Delta \Phi \approx \epsilon_{sat}\sqrt{2/3} \left(\sqrt{1+E_z z}-1\right)/E_z$, which shows that $\epsilon_N$ grows with propagation distance.

After the injection terminates, the injected electrons continue their betatron oscillations. For each injected electron $\Phi \gg 1$ and $\delta \Phi \ll \Phi $, $\Phi=\left(\sqrt{2\gamma} - \sqrt{2\gamma_0}\right) / E_z$ and $\gamma=\gamma_0+E_z z$ leading to $\delta \Phi/\Phi \approx \left( \delta z /z - \delta E_z/E_z\right)/2$. The variance of $\delta \Phi$ can be obtained by assuming the independence between the accelerating field and the injection time,
\begin{align}
\sigma_\Phi \equiv \sqrt{\left< \delta \Phi^2\right>} \approx \sqrt{\frac{1}{2E_z} \left[ \frac{\sigma_z^2}{z} + z \left( \frac{\sigma_{E_z}}{E_z}\right)^2\right]} \label{delta Phi}
\end{align}
where $\sigma_z^2=\left< \left(s_i -\left< s_i\right> \right)^2\right>$ and $\sigma_{E_z}^2=\left< \left( E_z - \left< E_z\right> \right)^2\right>$. To obtain an expression of $\Delta \Phi$ in terms of $\sigma_\Phi$, certain distribution of electrons needs to be assumed. For a uniform distribution, $\Delta \Phi = \sqrt{12}\sigma_\Phi$, Eq. (\ref{eq5}) then becomes 
\begin{align}
\epsilon_N = \epsilon_{sat} \sqrt{1- \left( \frac{\mathrm{sin} \sqrt{12}\sigma_\Phi}{\sqrt{12} \sigma_\Phi}\right)^2} \label{final emittance}
\end{align}
Eqs. (\ref{delta Phi}) and (\ref{final emittance}) predict that for $z <  \left(E_z / \sigma_{E_z} \right)  \sigma_z \equiv z_0$ the emittance actually decreases, it reaches a local minimum at $z_0$, and then increases until it saturates at $\epsilon_{sat}$. Therefore, to achieve the minimal emittance, the acceleration distance can be optimized to be close to $z_0$. 

\begin{figure}[bp]
\includegraphics[width=0.5\textwidth]{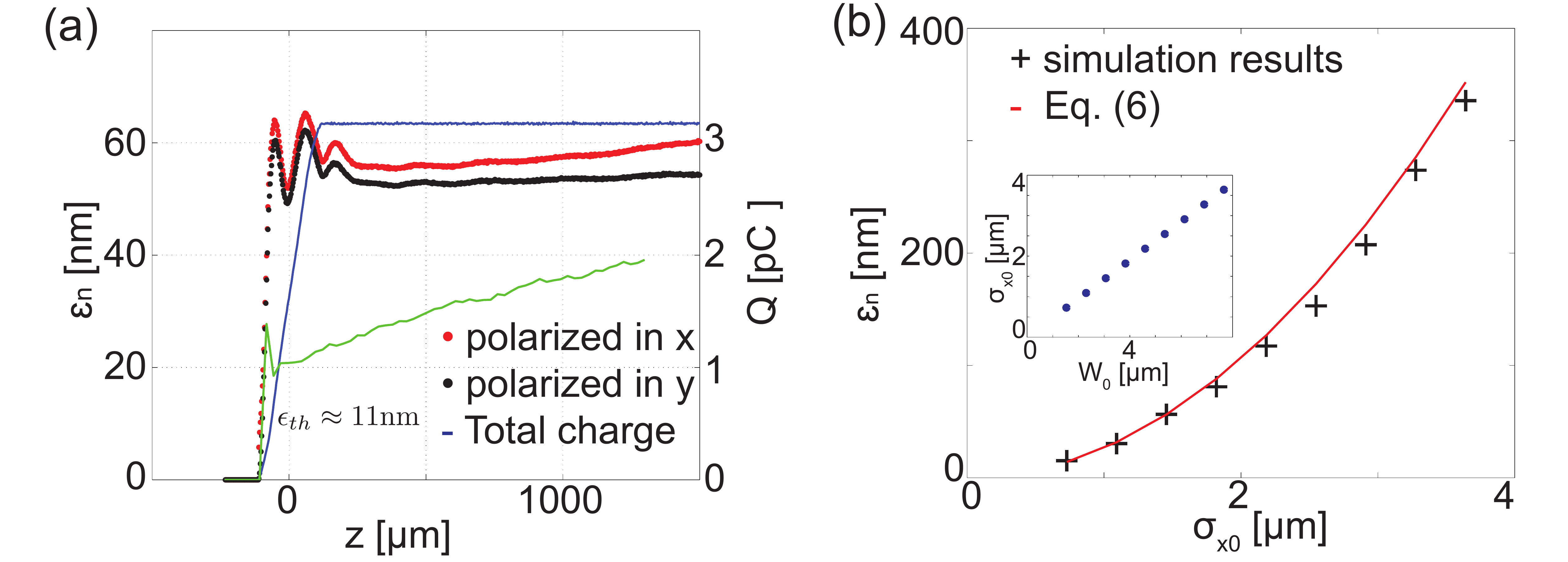}
\caption{\label{fig3} (a) Emittance evolution for large charge. The red and black lines show the emittance evolution for the laser in or out of the 2D simulation plane respectively. The green line shows the emittance evolution for limited Helium range ($19\micro\meter$), $n_{\mathrm{He}}=1.6\times 10^{17} \centi\meter^{-3}$. The charge value is obtained by assuming the beam would have been symmetric in a 3D simulation.  (b) The quadratic dependence of final emittances on $\sigma_{x0}$  for space charge dominated injection. The laser intensity was kept fixed at $a_0=0.04$ while scanning the spot size. The inner plot shows the dependence of $\sigma_{x0}$ on the laser spot size.}
\end{figure}

We next compare these predictions against self-consistent 2D OSIRIS simulations of laser triggered injection into the wake produced by an electron beam \cite{PhysRevLett.108.035001}. In the laser driver case, the evolution of phase space is sightly different in the directions parallel and perpendicular to the laser polarization direction due to the relatively large residual momentum in the polarization direction. In the direction perpendicular to the laser polarization, the evolution of phase space is almost identical to the beam driver case. In the polarization direction, the residual momentum makes the initial phase space distribution broader, however qualitatively the behavior is  very similar. 

As schematically  shown in Fig. \ref{fig2}(a), in the  simulation the beam driver propagates in a mixture of a fully ionized plasma of density $n_p=1.6\times 10^{17} \centi\meter^{-3}$ and a neutral He gas of density $n_{\mathrm{He}}=1.6\times 10^{13} \centi\meter^{-3}$. The simulation used a $5000\times4000$ cell grid and $2\times1$ and $2\times2$ particles per cell for  the plasma and neutral He respectively. In Fig. \ref{fig2}(b) we present the evolution of the emittance from the simulation as well as the predictions from Eq. (\ref{eq5}) (during ionization) and Eq. (\ref{final emittance}) (after ionization). We used $\sigma_z$ and $\sigma_{E_z}$ from simulations when plotting Eq. (\ref{final emittance}). The agreement is very good. The simulation curve also oscillates in time immediately after the rapid increase. In this case the ionization duration is limited by the Rayleigh length of the laser and is $\sim 0.6\pico\second > \pi \omega_p^{-1}$. Therefore the first group of electrons that are ionized will have rotated through more than an angle of $\pi$  in $x-p_x$ phase space. This is illustrated in Fig. \ref{fig1}(d) where the trajectory in $x-p_x$ space of an electron born at rest near $x=-1$ is shown. As it is accelerated its betatron frequency  and amplitude in $x$ decreases while its amplitude in $p_x$ increases. As a result the edge of the phase space for a collection of electrons is made up of layers of groups of electrons corresponding to each $\pi \omega_p^{-1}$ of injection. Each group corresponds to an ellipse with a different aspect ratios each of which have edges in phase space given by the trajectory shown in Fig. \ref{fig1}(d). The area in phase space will therefore oscillate at a frequency of twice the betatron frequency. After several oscillations the particles become smeared out in $x-p_x$ phase space and the oscillations damp away. This can also be seen in Fig. \ref{fig2}(c) ($x-p_x$ phase space from the simulation at the termination of the injection)  where the particles are not distributed in a simple ellipse. In Fig. \ref{fig2}(d) we show $x-p_x$ when $z=z_0$ (using the value of $\sigma_z,E_z,\sigma_{E_z}$ mentioned earlier). It is clearly seen that the range of $\Delta\Phi <\pi/2$ at this time even though $\Delta\Phi \approx 3\pi/2$ at the end of injection.

\begin{figure}[bp]
\includegraphics[width=0.5\textwidth]{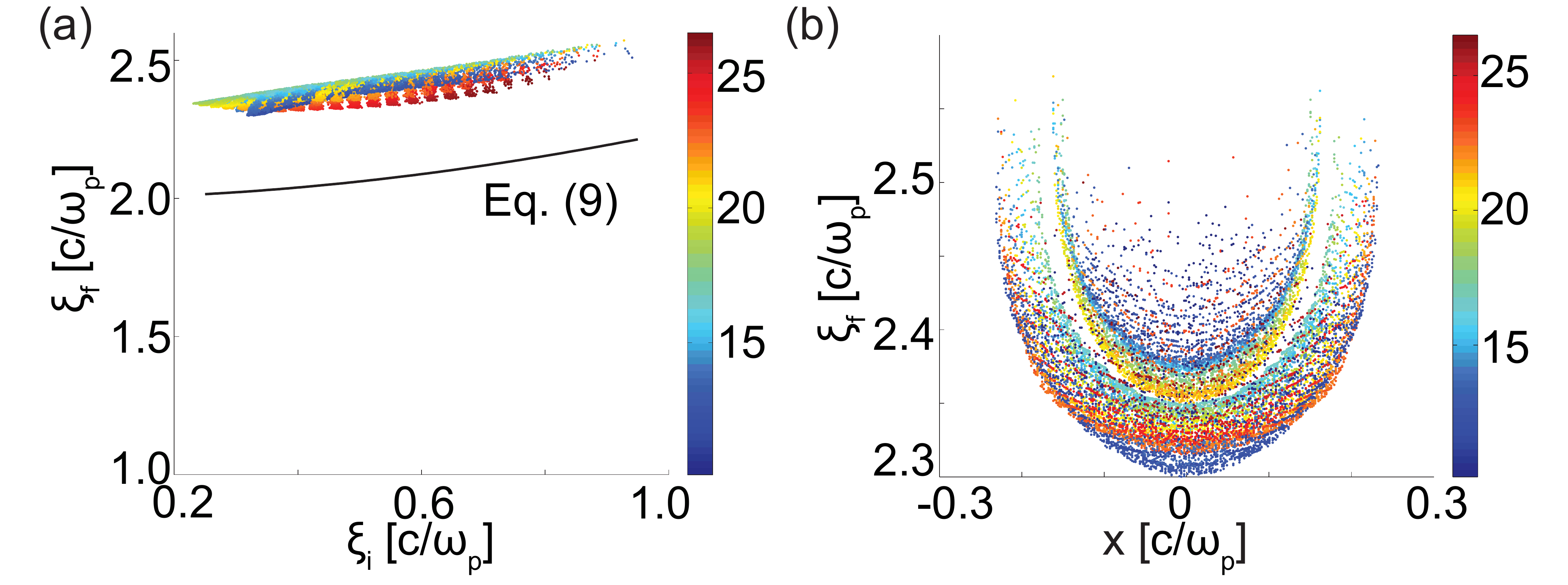}
\caption{\label{fig4}  (a) The dependence of $\xi_f$ on $\xi_i$ from simulation and the comparison with theoretical estimation. (b) The dependence of $\xi_f$ on $x_i$. Note: different colors represent different electron birth times.}
\end{figure}

Eqs. (\ref{delta Phi}) and (\ref{final emittance}) are based on Eq. (\ref{motion equation constant Ez}) where space charge forces are neglected. The relative importance of space charge is determined by the dimensionless quantity $F_{SC}/F_{acc}$ where $F_{SC}$ is the average space charge force and $F_{acc}$ is the acceleration force. In a typical photoinjector this ratio is much less than unity. However, in injection schemes using a plasma wakefield it can be much larger, where $F_{acc} \sim mc\omega_p$ and $F_{SC}/F_{acc} \sim n_b/\gamma^2 n_p > 1$ for $\gamma \sim 10$ ($n_b$ is the beam density, and typically $n_b / n_p \gtrsim 100$). The nonlinear space charge force of the injected electrons distorts the orbits of the electrons and fills out the available phase space area dictated by the initial conditions, leading to a saturation of the emittance around $\epsilon_{sat}$. In Fig. \ref{fig3}(a) this saturation behavior can be clearly seen. The quadratic dependence of $\epsilon_{sat}$ on the laser spot size is also verified through PIC simulations and very good agreement is achieved [Fig. \ref{fig3}(b)]. The theory and simulations show that for both the low and high charge regimes very small emittances can be achieved by limiting the ionization time to less than $\pi \omega_p^{-1}$. This can be seen in Eq. (\ref{eq5}) and in Fig. \ref{fig3}(a) where the green curve is from a simulation in which the ionization time was controlled by shortening the region of neutral He.

It turns out that the longitudinal mixing process also plays a critical role for the emittance dynamics. Due to this mixing, electrons ionized at different times can reside within the same longitudinal beam slice so that the projected and slice emittances have similar values. This is a fundamentally different situation comparing with the phase mixing process occurring in traditional accelerators \cite{PhysRevE.55.7565}. We analyze the longitudinal mixing by utilizing the injection threshold condition developed previously \cite{PhysRevLett.104.025003}. The injection condition can be approximated as $\delta \psi \approx -1$  where $\psi \equiv (e/mc^2)\left(\phi-v_\phi/c A_z \right)$ and  $\psi$ in the ion channel of the relativistic blowout regime can be expressed as $\psi(\xi,r) \approx  \left[ r_b^2 \left(\xi \right) - r^2 \right] / 4 $,  where $r_b(\xi)$ is the normalized radius of the ion channel (normalized to $k_p^{-1}$)
and it has a spherical shape for sufficiently large blowout radius $r_m$, i.e., $r_b^2(\xi)=r_m^2-\xi^2$ 
\cite{PhysRevLett.96.165002}\cite{lu2006nonlinearPoP}. In the beam driver case, we can obtain the final relative longitudinal position of each injected electron by applying $\delta \psi \approx -1$: 
\begin{align}
\xi_f \approx \sqrt{4+\xi_i^2} \label{final xi}
\end{align}
where the contributions from the initial and final radial position $r_i$ and $r_f$ are omitted since $r_i, r_f \ll1$. If the laser has a flattop intensity profile in the transverse dimensions, the initial positions  of the injected electrons $\xi_i$ depend only on their longitudinal positions $z_i$ and the birth times $t_i$, i.e., $\xi_i=ct_i - z_i+\xi_0(z)$, therefore each final slice of the injected beam is composed of the electrons ionized at different longitudinal position and different time. In real cases  where the laser intensity profiles are nonuniform and laser diffraction also plays a role, the birth time of each electron may vary due to the intensity variation. For example,  the electrons with larger transverse position $r_i$ are ionized with larger $\xi_i$ , and the electrons born before or after the laser focal point are ionized with larger $\xi_i$ than the electrons born near the focal point.

In Fig. \ref{fig4}(a) we plot the relation between $\xi_f$ and $\xi_i$ from a PIC simulation and compare with our theoretical estimate, and similar trend is obtained. In Fig. \ref{fig4}(b) we plot the relation between $\xi_f$ and initial $x$ of each electron, with each color representing a different birth time. One can see clearly that  the equal-time contours have U-shape, and this is mainly  due to the nonuniformity of the transverse Gaussian laser intensity profile. From the color code of both Figs. \ref{fig4}(a) and \ref{fig4}(b), the longitudinal mixing where electrons born at the same time are distributed into each slice is evident.


Work supported by NSFC grants 11175102, 11005063, thousand young talents program, DOE grants DE-FG02-92-ER40727, DE-SC0008491, DE-SC0008316, and NSF grants PHY-0936266, PHY-0960344. Simulations are performed on Hoffman and Dawson2 clusters at UCLA.

\bibliography{refs_xinlu}

\begin{thebibliography}{23}
\bibitem[{Xu2()}]{Xu2013prst1}
\bibinfo{note}{More detailed discussion will be presented in a future
  publication.}
\end{thebibliography}
\end{document}